\documentclass[dvips]{article}
\usepackage[english]{babel}
\usepackage[T1]{fontenc}
\usepackage[utf8]{inputenc}

\usepackage[switch]{lineno}

\usepackage{amsmath}
\usepackage{amssymb}
\usepackage{cite}
\usepackage{color}
\usepackage{subfigure}
\usepackage{url}
\usepackage{xspace}

\usepackage{icrc2011}

\title{Anomalous Longitudinal Shower Profiles and Hadronic Interactions}
\newcommand{\etal}{\MakeLowercase{\textit{et al. }}} 

\newcommand{\be}{\begin{equation}}
\newcommand{\ee}{\end{equation}}
\newcommand{\ben}{\begin{enumerate}}
\newcommand{\een}{\end{enumerate}}
\newcommand{\bi}{\begin{itemize}}
\newcommand{\ei}{\end{itemize}}
\newcommand{\bbe}{\begin{equation*}}
\newcommand{\eee}{\end{equation*}}
\newcommand{\bber}{\begin{equation*}\textcolor{red}}
\newcommand{\eeer}{\end{equation*}}
\newcommand{\bc}{\begin{center}}
\newcommand{\ec}{\end{center}}
\newcommand{\bea}{\begin{eqnarray}}
\newcommand{\eea}{\end{eqnarray}}
\newcommand{\bem}{\begin{pmatrix}}
\newcommand{\eem}{\end{pmatrix}}
\newcommand{\bbea}{\begin{eqnarray*}}
\newcommand{\eeea}{\end{eqnarray*}}

\newcommand{\xmax}{\ensuremath{X_\text{max}}\xspace}
\newcommand{\dx}{\ensuremath{\Delta X}\xspace}
\newcommand{\dxmax}{\ensuremath{\Delta X_\text{max}}\xspace}
\newcommand{\nmax}{\ensuremath{N_\text{max}}\xspace}

\newcommand{\ev}[1]{\ensuremath{10^{#1}\ \text{eV}}\xspace}
\newcommand{\gcm}{\ensuremath{\text{g}~\text{cm}^{-2}}\xspace}

\newcommand{\chis}{\ensuremath{\chi^2}\xspace}

\newcommand{\conex}{{\scshape conex}\xspace}
\newcommand{\qgsjet}{{\scshape qgsjetII}\xspace}
\newcommand{\epos}{{\scshape epos}\xspace}
\newcommand{\sibyll}{{\scshape sibyll}\xspace}

\shorttitle{C.\ Baus \etal Anomalous Longitudinal Shower Profiles and
  Hadronic Interactions}
\authors{C.~Baus$^{1}$, R.~Engel$^{1}$, T.~Pierog$^{1}$, R.~Ulrich$^{1}$, M.~Unger$^{1}$}
\afiliations{$^{1}$Karlsruhe Institute of Technology, Germany}
\email{colin.baus@student.kit.edu}
\abstract{ The bulk of air showers initiated by very high energy
  cosmic rays exhibits a longitudinal development in depth with a single
  well-defined shower maximum. However, a small fraction of showers has
  a profile that differs considerably from this
  average behaviour. In extreme cases, such anomalous longitudinal
  profiles can even have two distinct shower maxima. We discuss the
  properties of the primary interactions that lead to such
  profiles. Simulations are used to estimate the rate of anomalous
  profiles in dependence of primary energy, mass, and characteristic
  features of hadronic multiparticle production at very high energies. }

\begin{document}
\maketitle
\section{Introduction}
Cosmic ray detectors like HiRes~\cite{AbuZayyad:2000uu}, the Pierre Auger Observatory~\cite{Abraham:2004dt}, or
Telescope~Array~\cite{telescopearray} measure the longitudinal
profiles of air showers initiated by ultra-high energy
cosmic ray particles. The data of these experiments offers
 the possibility to study hadronic interactions beyond the reach of man-made
accelerators. In practice, however, the interpretation of the data is difficult
because of the unknown mass composition of the cosmic particle beam.

In this paper, we present a Monte Carlo study of longitudinal profiles using
the \conex air shower simulation program~\cite{Bergmann:2006yz} to study if
'anomalous' longitudinal shower profiles could provide a unique signature
for properties of hadronic interactions.
Measurements~\cite{AbuZayyad:2000np}, simulations~\cite{unversalprofile} and
theoretical considerations (e.g. \cite{Lipari:2008td}) suggest
that the longitudinal development of the electromagnetic component of
ultra-high energy air showers is to a good approximation universal, i.e. that
the shape of the profiles is very similar for the bulk of showers irrespective
of the primary energy, particle type or hadronic interaction model.
In rare cases, however, simulated profiles can significantly deviate
 from this universal shape and {\itshape anomalous} shower profiles
can be found that in extreme cases show two completely separated maxima.

\begin{figure*}[!t]
\centering
\subfigure[Helium $E_0=\ev{17}$ (\epos)]
{
\includegraphics[width=0.32\textwidth]{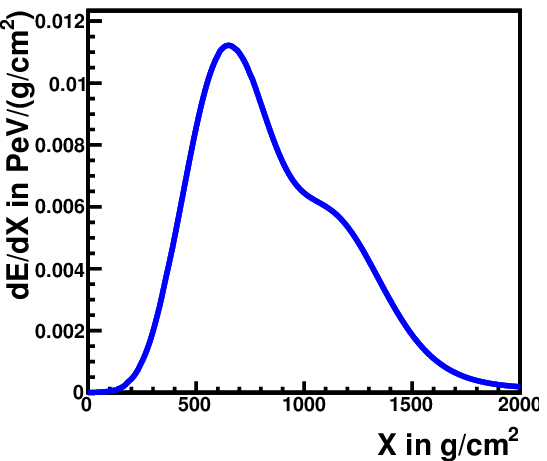}\hfill
\label{fig:motiexmpl}
}
\subfigure[Proton $E_0=\ev{19.5}$ (\sibyll)]
{
\includegraphics[width=0.32\textwidth]{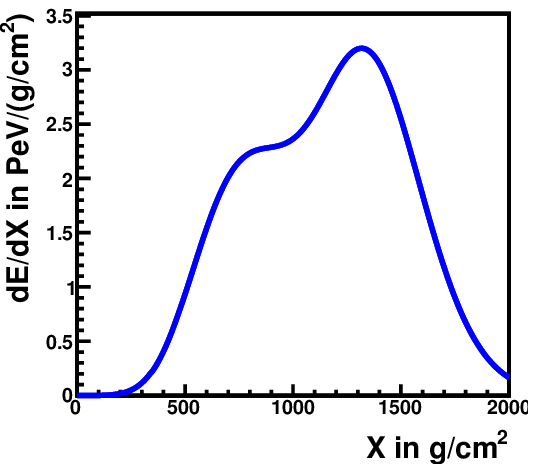}\hfill
}
\subfigure[Proton $E_0=\ev{20}$ (\sibyll)]
{
\includegraphics[width=0.32\textwidth]{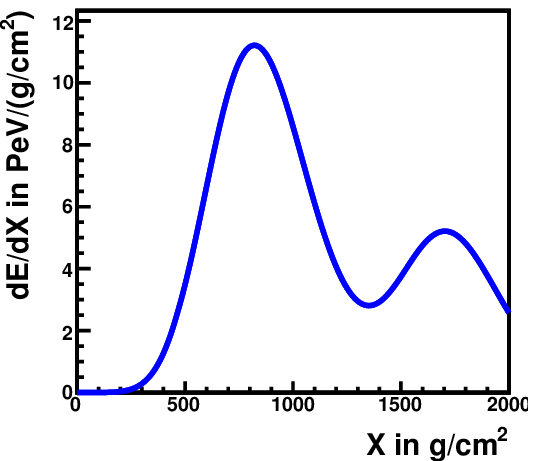}
}
\caption{Examples of anomalous shower profiles simulated with \conex.}
\label{fig:motivation}
\end{figure*}

Examples of such anomalous shower profiles are shown in Fig.~\ref{fig:motivation}.
The profile in the left panel is generated by a primary helium nucleus
with an energy of \ev{17}. The first interaction occurs at $19\ \gcm$
and results in a sub-shower with a maximum at around $700\ \gcm$.
A spectator nucleon with a quarter of the primary energy penetrates
deeply into the atmosphere before it interacts at $583\ \gcm$ and creates
a second sub-shower that reaches its maximum beyond $1000 \gcm$.
Similar cases are shown in Fig.~\ref{fig:motivation}(b) and (c),
but now for primary protons. Here, instead of a spectator nucleon, the
second shower is created by the leading particle from the first interaction.

These examples suggest, that the anomalous
shower profiles originate from leading or spectator particles that penetrate
deeply into the atmosphere before interacting. The probability for
propagating a slant depth distance greater than $\dx$ without
interacting is given by
\be
P\left(\dx\right) = e^{-\frac{\dx}{\lambda}},
\label{eq:int}
\ee where $\lambda$ denotes the hadronic interaction length in air.
Assuming that a propagation distance of at least $300\ \gcm$ is needed
to experimentally distinguish these showers from the universal shower
profile, one would expect $P\simeq 3\cdot10^{-3}$ and $P\simeq
8\cdot10^{-4}$ for \ev{17} and \ev{19} primaries using a hadronic
interaction length as predicted by the \sibyll interaction model. The
actual rate of occurrence will however be smaller, because for
an experimental observation of anomalous shower profiles,
the inelasticity of the interaction
must be in a suitable range such that both sub-showers carry a
fraction of the primary energy that is large enough for a detection.


\section{Analysis of Air Shower Simulations}
To estimate the frequency of detectable anomalous air showers we
generated more than $10^{4}$ showers for each combination of
three different hadronic
interaction models ({\scshape Sibyll}2.1~\cite{sibyll}, {\scshape
  Qgsjet}II~\cite{qgsjet} and {\scshape Epos}1.99~\cite{epos}), seven
primary energies between $E= 10^{17}$ and $10^{20}$~eV and three
primary masses (proton, helium, and iron).

Anomalous profiles are identified by comparing the $\chi^{2}$-value
 obtained from a fit of the simulated profile with a Gaisser-Hillas
function~\cite{1977ICRC....8..353G}, $f_\text{GH}$,
with the $\chi^{2}$ that results from
the sum of two Gaisser-Hillas functions:
\begin{align}
  \chis_\text{double}&=\sum\limits_{i=0}^{n}\frac{\left(N_i-f_\text{GH1}(X_i)-f_\text{GH2}(X_i)\right)^2}{V_i},
\label{chi2}
\end{align}
with
\begin{align}
  f_\text{GH1}&=f_\text{GH}\left({\nmax}_{,1}, {\xmax}_{,1},\lambda_1, X_0^1\right)\\
  f_\text{GH2}&=f_\text{GH}\left({\nmax}_{,2}, \dxmax,\lambda_2, X_0^2\right),
\end{align}
where $N_i$ denotes the simulated shower size at slant
depth $X_i$ and $n$ is the number of generated data points.
The normalisation of the functions is given by ${\nmax}_{,i}$
and ${\xmax}_{,1}$ is the depth of maximum of the first sub-shower.
Following~\cite{showerrec}, the shape parameters
$\lambda_{i}$ and $X_0^{i}$ are constrained to average values corresponding to
the 'normal' universal profiles to assure a better stability of the fit.
Finally, \dxmax denotes the difference between the shower maxima of the
first and second sub-shower.

For an anomalous profile that originates
from a deeply penetrating leading particle, we expect that
the fit with two
Gaisser-Hillas functions will lead to parameters that are
related to the properties of the two sub-showers.
The inelasticity $\kappa$ of the first interaction
can be estimated from the normalisations of the two Gaisser-Hillas functions,
\begin{align}
  \kappa & \approx 1-\frac{{\nmax}_{,2}}{{\nmax}_{,1}+{\nmax}_{,2}},
\label{eq:kappafitted}
\end{align}
and the fitted distance between the two shower maxima is related to
the slant depth distance between the first and second interaction
\begin{align}
  \dx & \approx \dxmax\label{eq:dxfitted}.
\end{align}
This relations assumes that the shower development from the
interaction point is a mere translation, that is the same for both
showers. This should be a good approximation, since the average slant
depth distance between the interaction point and the shower maximum
depends only logarithmically on energy.

In order to obtain a meaningful \chis in Eq.~(\ref{chi2}), a choice must be
made for $V_i$.
\conex generates 'exact' shower sizes without statistical fluctuations.
A Poissonian ansatz, $V_i = N_i$ would lead to an increasing
sensitivity of the $\chi^2$s to deviations from a universal profile
 with energy. For this study however, we prefer to have a constant
sensitivity at each energy and therefore use $V_i = k N_i/E$,
where $k$ is chosen such that $\sqrt{\sum V_i} / \sum N_i = 0.01$.
With this re-normalisation, equal \chis
probabilities for all energies are assured that enable us to
study the 'intrinsic' evolution of the fraction of anomalous profiles
with energy.
Furthermore, profile points with $\sqrt{V_i} / N_i > 0.3$ are excluded from the fit
to avoid the influence of the long tail of deeply penetrating muonic
component of the shower size profile that can not be well described
by a four parameter Gaisser-Hillas function; Fluorescence detectors
can not observe this tail above the night-sky background.\\

For the analysis of the generated showers, each profile point is fluctuated
by $\sqrt{V_i}$ using Gaussian random numbers and the $\Delta \chi^2$
between single- and double-Gaisser-Hillas fit is calculated. Then we define
a shower as 'anomalous' if
\begin{itemize}
  \item two Gaisser-Hillas functions significantly improve
         the goodness of fit ($\Delta \chi^2 > 25$),
  \item the shower maxima are clearly separated ($\dxmax > 300\ \gcm$),
  \item both fitted \xmax were found within the profile range,
  \item both fitted sub-showers have more than 20\% of the primary energy.
\end{itemize}
\section{Results}
\begin{figure*}[t!]
\centering
\subfigure[\sibyll]
{
\includegraphics[width=0.315\linewidth]{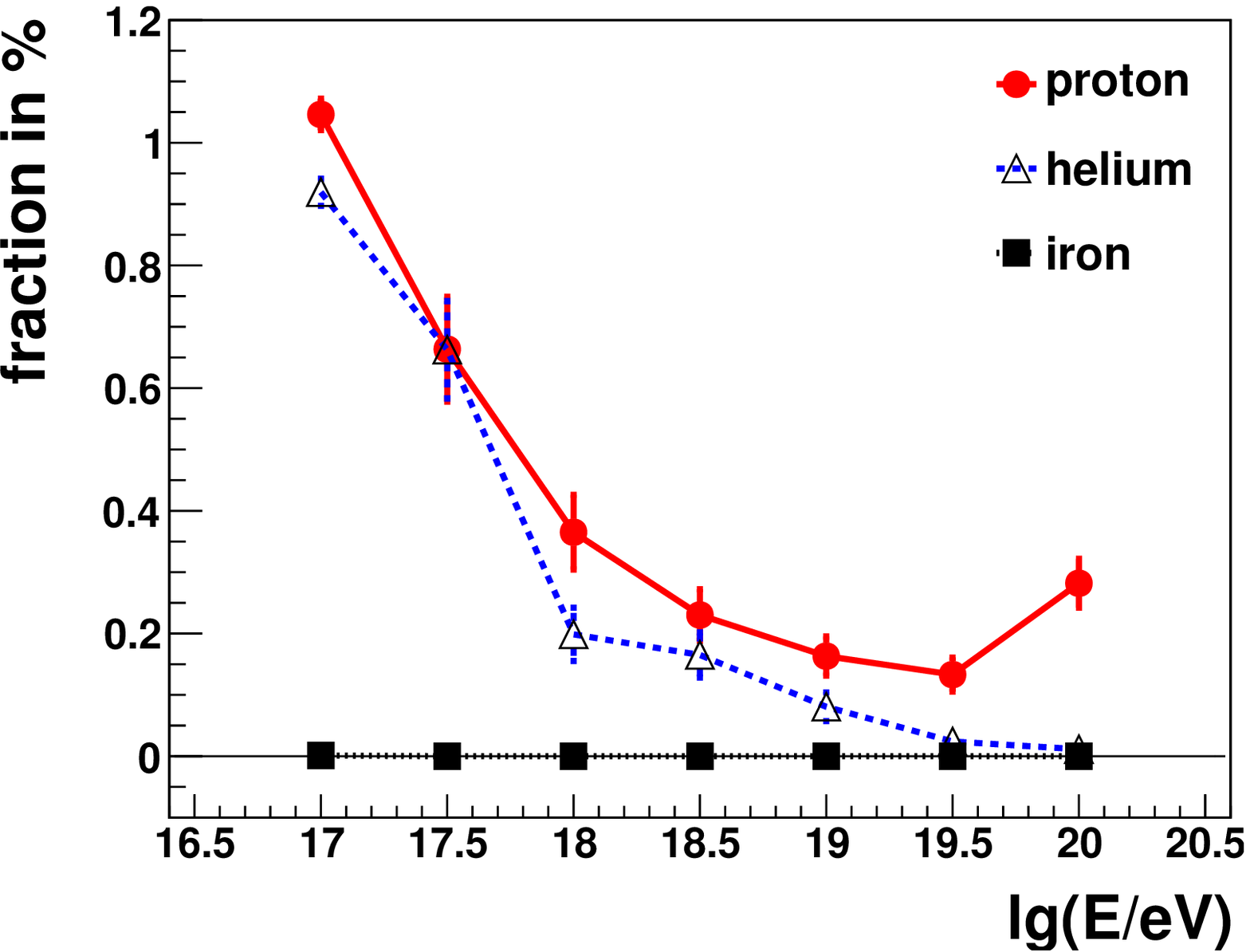}
\label{fig:dbfracsib}
}
\hfill
\subfigure[\qgsjet]
{
\includegraphics[width=0.315\linewidth]{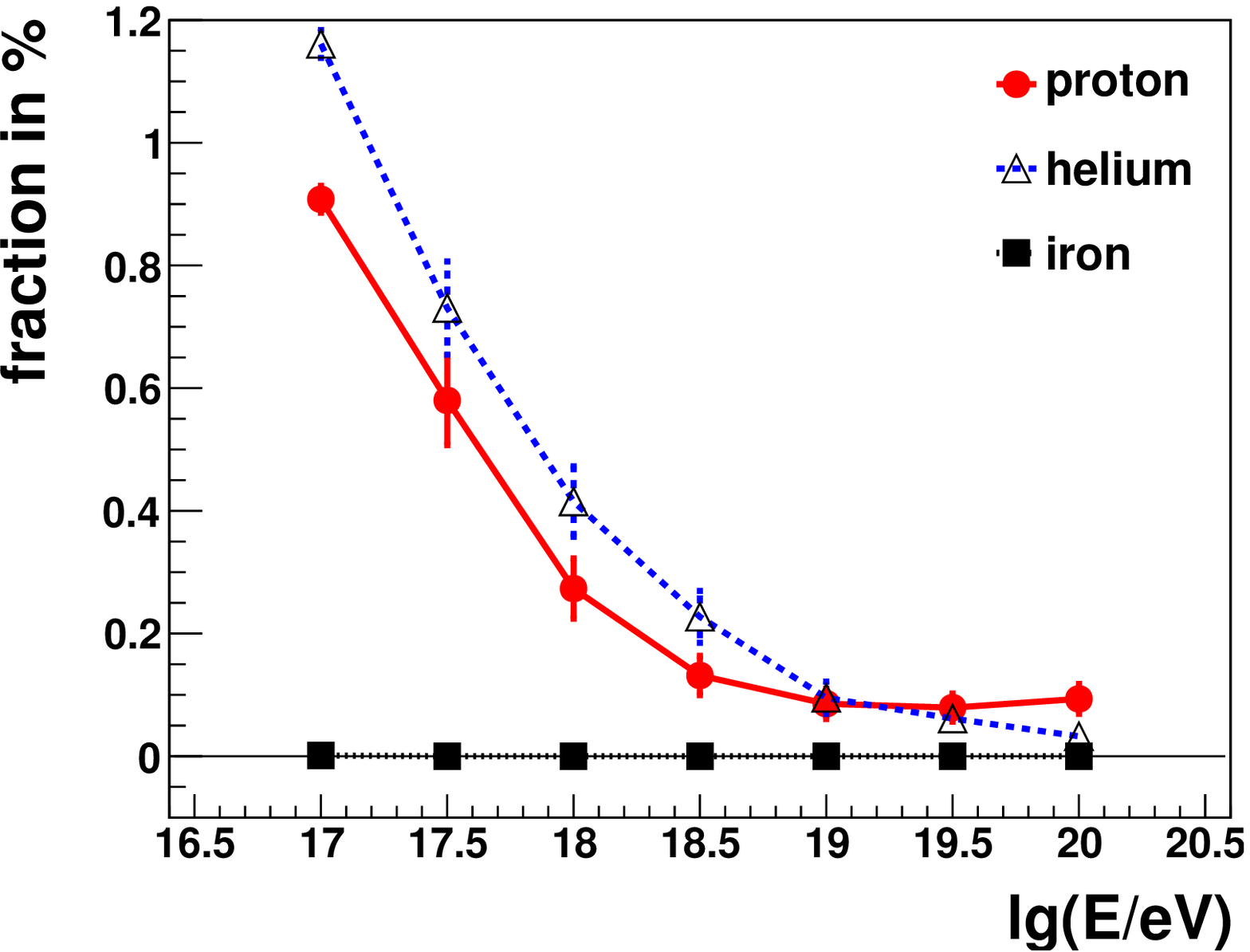}
}
\hfill
\subfigure[\epos]
{
\includegraphics[width=0.315\linewidth]{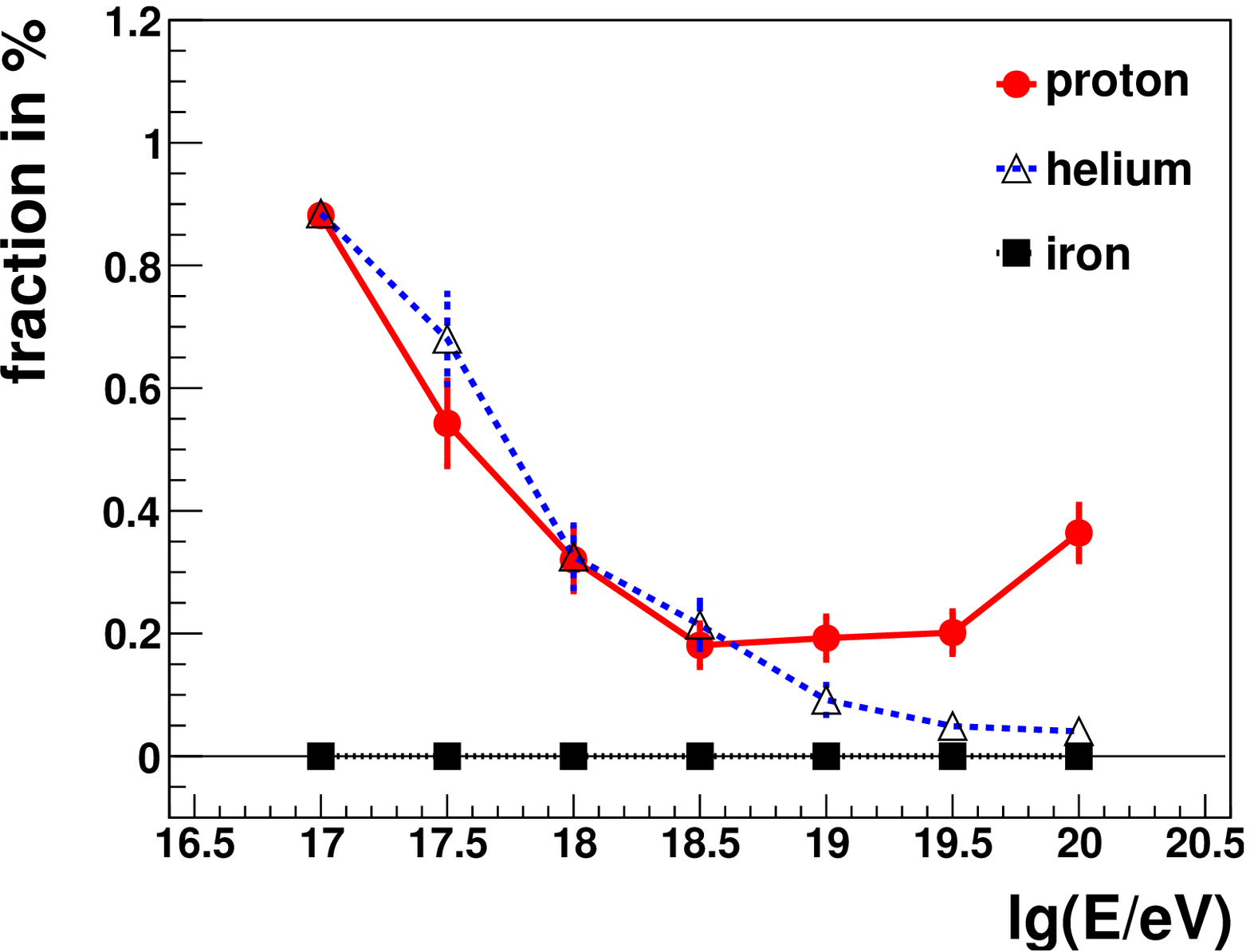}
}
\caption{Fraction of anomalous shower profiles for different primaries and
  interaction models.}
\label{fig:dbfrac}
\end{figure*}

Using the criteria from the last section to select and define anomalous
shower profiles, their fraction of the total simulated sample can be
determined. As can be seen in Fig.~\ref{fig:dbfrac}, no anomalous
showers are predicted for iron nuclei, which is not surprising because
spectator nucleons can carry less than 2\% (1/56) of the primary energy.
Heavier fragment nuclei could have a larger energy, but due to their large
cross section they have a small mean free path length in the atmosphere.

Anomalous shower profiles of helium and proton primaries occur with a
very similar frequency and the fractions are of the same order of
magnitude as estimated with the simple analytic model,
Eq.~\ref{eq:int}. The fraction decreases with energy as it would be
expected due to the rise of the cross sections with energy in the
simulations. In case of \sibyll and \epos, the fraction rises
slightly again at the highest energies. The reason for this behaviour
could be the Landau-Pomeranchuk-Migdal (LPM) effect~\cite{LPM} on photons produced by the decay of a leading $\pi^0$.
The LPM effect significantly increases the mean free path of electromagnetic
particles above $10^{18}-10^{19}$~eV, and in the case of \sibyll and \epos a
leading $\pi^0$ can carry more than 50\% of the primary energy. In the case
of \qgsjet no leading $\pi^0$ are produced. Further studies are needed to confirm
this hypothesis.\\

\begin{figure}[b!]
\centering
\includegraphics[width=\linewidth]{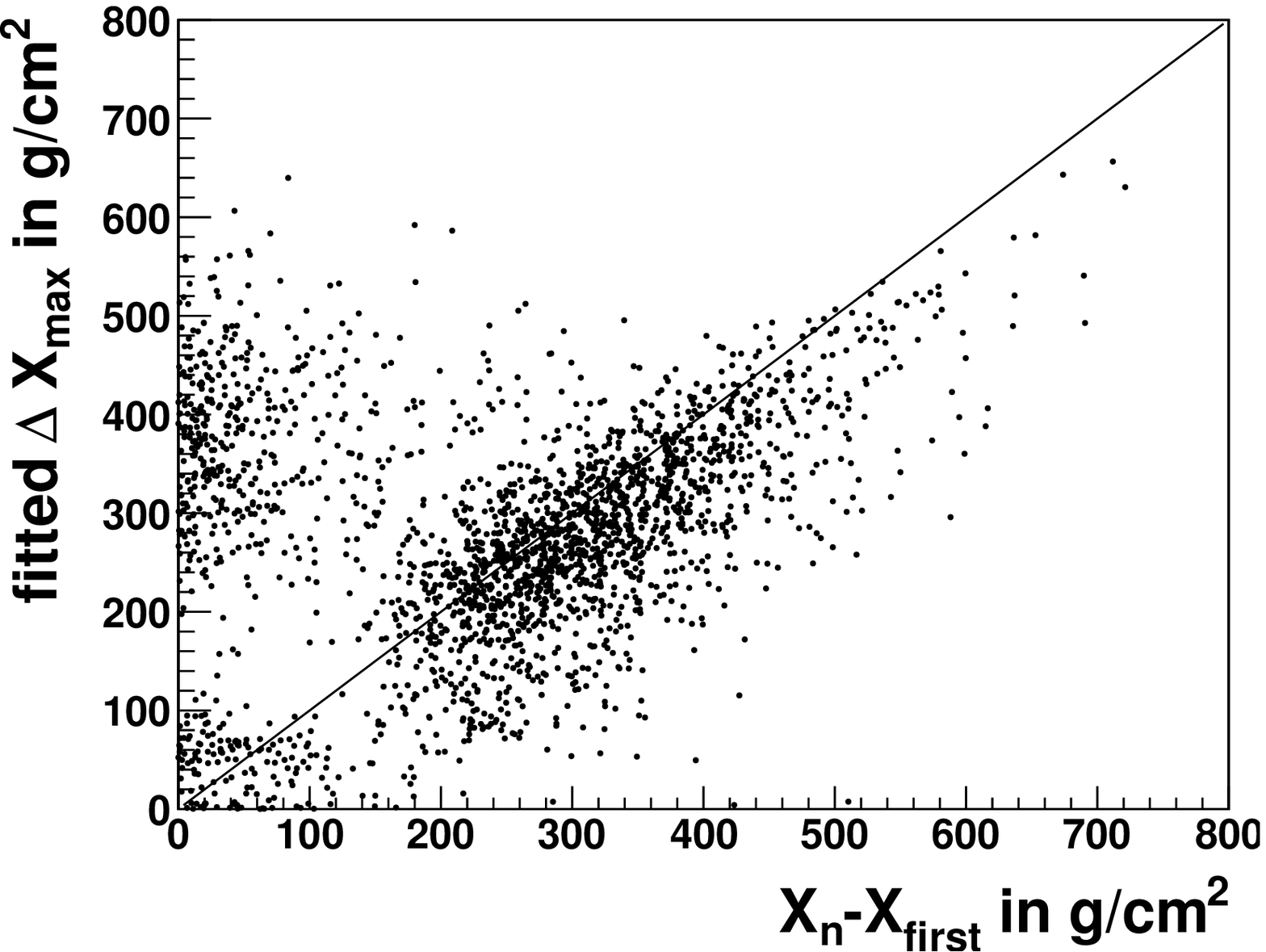}
\caption{Correlation between \dx and \dxmax for protons (\sibyll).}
\label{fig:corr}
\end{figure}

The correlation between the fitted \dxmax and the slant depth
difference $\dx=X_{n}-X_{\rm first}$ of the two most relevant
interaction points is shown in Fig.~\ref{fig:corr}. Here $X_{\rm
  first}$ is the interaction depth of the primary particle in the
atmosphere and $X_{\rm n}$ the interaction depth of the $n$th leading
particle\footnote{For $n=1$ this refers to the leading particle of the
  primary interaction, for $n=2$ to the leading particle of the
  interaction of the leading particle of the primary interaction,
  etc..}, where interactions are only considered if they have a
minimal inelasticity of $\kappa>0.15$.  As can be seen, the
distance between the shower maxima is a good estimator for $\dx$ for
the majority of the showers.  In addition to the events populating the
diagonal, there is a cluster of events at small \dx but large
\dxmax. These events are due to deeply penetrating sub-showers created
\emph{not} by leading particles, so $X_{n}$ and thus also \dx are not the correct
quantities.

\begin{figure}[t!]
\centering
\includegraphics[width=\linewidth]{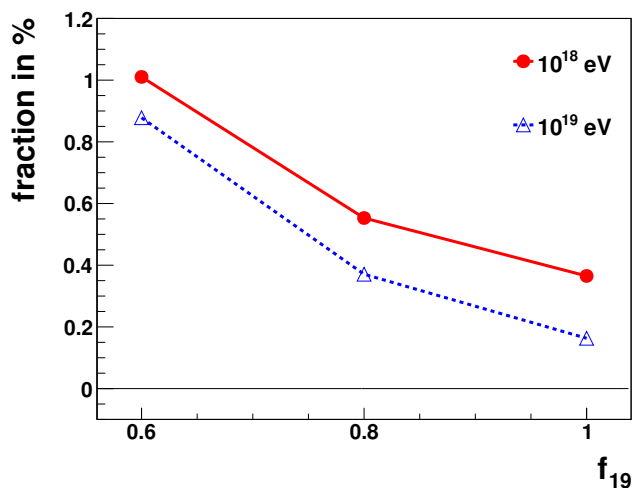}
\caption{Anomalous profile fraction for a modified cross section in
  \sibyll.}
\label{fig:dbfracmod}
\end{figure}

As a further test of the sensitivity of the anomalous profile fraction
to the hadronic interaction length, the response of the event
fraction to a decrease of the hadronic cross sections in the simulation
is studied.
For this purpose all cross section above \ev{15} are rescaled
logarithmically during \conex simulations by a factor $f(E)$,
\be
f(E)=1+\left(f_{19}-1\right)\frac{\text{ln}\left(E/\ev{15}\right)}{\text{ln}(\left(\ev{19}/\ev{15}\right)},
\label{eq:mod}
\ee
where $E$ is the energy and $f_{19}$ is the rescaling factor of the
cross section at \ev{19}~\cite{Ulrich:2010rg}.
The resulting fractions for diminished cross sections are shown in
Fig.~\ref{fig:dbfracmod}. As can be seen, decreasing the cross-section of
\sibyll to $80\%$ means a $50\%$ and $100\%$ higher fraction at
\ev{18} and \ev{19} respectively and a reduction to $60\%$
increases the rate even further.

\section{Conclusions}
In this paper we presented a simulation study on anomalous
longitudinal shower profiles. These showers exhibit in extreme cases
two distinct shower maxima and originate from deeply penetrating
spectator nucleons or leading particles. It was shown that
their rate is expected to be
largest at low energies and for light primary masses. It is however
worthwhile noting that the absolute rate estimates given in
Fig.~\ref{fig:dbfrac} are only indicative and depend on the selection
criteria as well as on the precision of the profiles, $V_i$. A
precise estimate of the rate of anomalous showers detectable by current
detectors requires to study detailed detector simulations, which is beyond the
scope of this work.

The experimental detection of such showers would be an unambiguous
proof of the presence of light nuclei in the cosmic particle beam. Moreover,
if it is possible to detect enough of these events, they could provide a novel tool
to study the cross section and inelasticity of hadronic interactions at ultra-high
energies.

\bibliographystyle{unsrt} 

\clearpage
\end{document}